\documentclass[useAMS,usenatbib]{mn2e}

\usepackage{epsfig}
\usepackage{longtable}
\usepackage{times}

\newcommand{\sw}{$Swift$}

\def \inte {\emph{INTEGRAL}}
\def \xmm {\emph{XMM--Newton}}

\def \sw {{\it Swift}}
\def \rxte {\emph{RXTE}}
\def \src {\mbox{IGR~J16418$-$4532}}

\def \hcm {\hbox {\ifmmode $ atom cm$^{-2}\else atom cm$^{-2}$\fi}}

\def \arcsec {\hbox{$^{\prime\prime}$}}

\def \ATel {Astron.\ Tel.}
\def \apj {ApJ}
\def \apjl {ApJL}

\def \aj {AJ}
\def \aap {A\&A}

\def \mnras {MNRAS}

\title[IGR~J16418$-$4532 with Swift]{\emph{Swift}/XRT monitoring of the candidate 
Supergiant Fast X--ray Transient IGR~J16418$-$4532}

\author[P.\ Romano et al.]{P.\ Romano$^{1}$,  
       V.\ Mangano$^{1}$, L.~Ducci$^{2}$, P.\ Esposito$^{3}$, P.A. Evans$^{4}$, S.\ Vercellone$^{1}$, 
\newauthor  J.A.~Kennea$^{5}$, D.N.~Burrows$^{5}$, N.~Gehrels$^{6}$  \\  
$^{1}$INAF, Istituto di Astrofisica Spaziale e Fisica Cosmica,
        Via U.\ La Malfa 153, I-90146 Palermo, Italy\\
$^{2}$  Institut f\"ur Astronomie und Astrophysik,
         Universit\"at T\"ubingen, Sand 1, D-72076 T\"ubingen, Germany \\
$^{3}$INAF, Osservatorio Astronomico di Cagliari, localit\`a Poggio dei Pini, 
     Strada 54, I-09012 Capoterra, Italy\\ 
$^{4}$Department of Physics \& Astronomy, University of Leicester, LE1 7RH, UK\\
$^{5}$Department of Astronomy and Astrophysics, Pennsylvania State 
        University, University Park, PA 16802, USA\\
$^{6}$NASA/Goddard Space Flight Center, Greenbelt, MD 20771, USA
\\
}

\begin{document}

\date{Accepted 2011 September 28.  Received 2011 September 28; in original form 2011 August 17}

\pagerange{\pageref{firstpage}--\pageref{lastpage}} \pubyear{2011}

\maketitle

\label{firstpage}

\begin{abstract}
{We report on the {\it Swift} monitoring of the candidate supergiant 
fast X-ray transient (SFXT) IGR~J16418$-$4532, for which both orbital and 
spin periods are known ($\sim 3.7$\,d and $\sim 1250$\,s, respectively). 
Our observations, for a total of $\sim 43$\,ks,  
span over three orbital periods and represent 
the most intense and complete sampling of the light curve of 
this source with a sensitive X-ray instrument. 
With this unique set of observations we can address the nature of this transient. 
By applying the clumpy wind model for blue supergiants to the observed 
X-ray light curve,
and assuming a circular orbit, 
the X-ray emission from this source can be 
explained in terms of the accretion from a spherically symmetric clumpy wind, 
composed of clumps with different masses, ranging from 
$\sim5\times10^{16}$~g to $10^{21}$~g. 
Our data suggest, based on the X-ray behaviour, that this is 
an intermediate SFXT. 
}

\end{abstract}



	\section{Introduction\label{igr16418:intro}}

%
The X-ray transient \src\ was discovered by \inte\ during observations 
of the black hole X-ray transient 4U~1630-47 on 2003 February 1--5
\citep{Tomsick2004:atel224}, 
with a 20--40\,keV flux of $3\times10^{-11}$ erg cm$^{-2}$ s$^{-1}$. 
\citet{Sguera2006} reported on further \inte\ observations, taken 
on 2004 February 26, during which fast ($\sim 1$\,hr) X-ray outbursts 
were observed that peaked at $\sim 80$\,mCrab (20--30\,keV). 
This behavior led \citet{Sguera2006} to propose that \src\ is a 
candidate supergiant fast X-ray transient (SFXT). 
SFXTs \citep[e.g.\ ][]{Sguera2005} are high mass X-ray binaries (HMXBs) 
associated with OB supergiant stars via optical spectroscopy that
show an X-ray dynamic range of 3--5 orders of magnitude.
Indeed, they are characterized by bright (peak luminosities of 10$^{36}$--10$^{37}$~erg~s$^{-1}$), 
short (a few hours, as observed by \inte, \citealt{Sguera2005,Negueruela2006:ESASP604})
X-ray outbursts significantly shorter than those of typical Be/X-ray binaries, 
and a quiescent luminosity of  $\sim 10^{32}$~erg~s$^{-1}$ 
\citep[e.g.\ ][]{zand2005,Bozzo2010:quiesc1739n08408}.

In their systematic analysis of the \inte\ observations from 2003 to 2009, 
\citet{Ducci2010} detected 23 outbursts, for an activity duty cycle of 
$\sim 1$\,\% (one of the highest among the 14 SFXTs and SFXT candidates 
they examined) 
and fluxes ranging between $1.3\times10^{-10}$ erg cm$^{-2}$ s$^{-1}$ 
and $4.8\times10^{-10}$ erg cm$^{-2}$ s$^{-1}$ (18--100\,keV). 
\xmm\ observations \citep{Walter2006} showed a 
heavily absorbed ($N_{\rm H}=(1.0\pm0.1)\times10^{23}$ cm$^{-2}$) 
pulsar ($P_{\rm spin}=1246\pm100$\,s) with an unabsorbed flux of
$1.3\times10^{-10}$ erg cm$^{-2}$ s$^{-1}$  (2--100\,keV) in a fit
performed in conjunction with the \inte\ data.

\citet{Chaty2008} proposed, based on the \xmm\ position,  
2MASS~J16415078$-$4532253 as the best NIR counterpart candidate; 
they also fitted the spectral energy distribution (SED), obtaining a 
temperature of 32\,800\,K for the massive companion, consistent with 
an OB spectral classification. 
SED fitting by \citet{Rahoui2008} yielded $A_{V}=14.5$\,mag, 
and a O8.5 spectral type; depending on the luminosity class 
they derive a distance of 4.9 (V), 8.3 (III), or 13\,kpc (I). 

A periodicity of $\sim 3.75$\,d was discovered by \citet{Corbet2006:atel779}
in the \sw/Burst Alert Telescope   \citep[BAT, ][]{Barthelmy2005:BATmn}  ($3.753\pm0.004$\,d, MJD 53360--53630) and 
\rxte/ASM ($3.7389\pm0.0004$\,d, MJD 50091--53810)
light curves, 
which was interpreted as the orbital period of a binary system. 
The hint of a total eclipse at phase 0.5 observed in the BAT data 
is consistent with a supergiant companion. 
\citet{Levine2010} refined the orbital period value to $3.73886\pm0.00028$\,d, 
by using the \rxte/ASM data up to MJD 55224. 
The observed values of the orbital and pulse periods indeed place \src\
in a region of the Corbet diagram \citep{Corbet1986}  
where wind-accreting  HMXBs generally lie.  

In this paper we analyze all the \sw\ data collected on \src. 
The first set of observations comprises two  
2\,ks target of opportunity (ToO) observations performed on
2011 February 18 and March 4, following a MAXI alert of a possible detection 
of \src\ on 2011 February 17 \citep[preliminary results were described in ][]{Romano2011:atel3174}. 
The second set is drawn from a monitoring campaign  in 2011 July that spans over 
three orbital periods. 
This \sw\ monitoring represents the most intense and complete sampling along the orbital period
of the light curve of this source with a sensitive X-ray instrument.

 \begin{table}
 \begin{center}
\tabcolsep 3pt   
 \caption{Summary of the {\it Swift}/XRT observations.\label{igr16418:tab:xrtobs} }
 \begin{tabular}{lllll}
 \hline
 \noalign{\smallskip}
 ObsID     & Start time  (UT)  & End time   (UT) & Exp. & $\phi^a$   \\ 
              &   &  &(s)        \\
  \noalign{\smallskip}
 \hline
 \noalign{\smallskip}
00031929001$^b$     &2011-02-18 01:59:43  &2011-02-18 03:43:26 & 1951 & 0.27  \\ 
00031929002     &2011-03-04 00:03:19  &2011-03-04 05:01:57 & 2001 & 0.01  \\
00043105001$^c$ &2011-06-27 05:14:54  &2011-06-27 05:23:56 & 537  & 0.80  \\
00031929003     &2011-07-13 09:46:55  &2011-07-13 14:43:57 & 4756 & 0.16  \\
00031929004     &2011-07-14 12:39:39  &2011-07-14 19:31:56 & 4832 & 0.47  \\
00031929005$^d$ &2011-07-15 06:42:44  &2011-07-15 23:00:58 & 273  & 0.72  \\
00031929005     &2011-07-15 06:42:50  &2011-07-15 23:01:57 & 1078 & 0.72  \\
00031929006$^d$ &2011-07-16 10:15:02  &2011-07-16 23:05:36 & 458  & 0.01  \\
00031929006     &2011-07-16 10:15:05  &2011-07-16 23:06:58 & 379  & 0.01  \\
00032037001     &2011-07-18 21:28:28  &2011-07-19 00:51:31 & 1947 & 0.62  \\
00031929007     &2011-07-19 01:52:41  &2011-07-19 23:19:57 & 2673 & 0.77  \\
00032037002     &2011-07-20 05:18:42  &2011-07-20 15:22:58 & 3483 & 0.01  \\
00032037003     &2011-07-21 10:06:18  &2011-07-21 15:27:54 & 3761 & 0.31  \\ 
00032037004     &2011-07-22 15:19:43  &2011-07-22 18:44:57 & 2812 & 0.62  \\
00032037005     &2011-07-23 11:48:19  &2011-07-23 15:23:56 & 4127 & 0.85  \\  
00032037006     &2011-07-29 12:13:30  &2011-07-29 23:34:57 & 4650 & 0.50  \\
00032037007     &2011-07-30 01:01:30  &2011-07-30 09:27:56 & 4287 & 0.63  \\
  \noalign{\smallskip} 
  \hline
  \end{tabular}
  \end{center}
  \begin{list}{}{} 
  \item[$^{\mathrm{a}}$ Mean phase referred to 
    $T_{\rm epoch}=$ MJD 53560.2 \citep{Levine2010}.] 
  \item[$^{\mathrm{b}}$ Corresponding to MJD 55610.08.] 
  \item[$^{\mathrm{c}}$ Serendipitous observation.] 
  \item[$^{\mathrm{d}}$ WT mode data.] 
  \end{list}   
\end{table}

 	 \section{Observations and Data Reduction\label{igr16418:dataredu}}

Table~\ref{igr16418:tab:xrtobs} reports the 
log of the \sw/X-ray Telescope  \citep[XRT, ][]{Burrows2005:XRTmn}  observations used in this paper. 
The \sw\ observations of \src\ during 2011 February and March
(00031929001 and 00031929002) were obtained \citep{Romano2011:atel3174}
as two ToO observations following a MAXI alert (5609524433, later retracted, 
as due to GX~340$+$0) of a possible detection of \src\ on 2011 February 17. 
The data collected in 2011 July were obtained as a ToO 
monitoring program of 9 scheduled daily observations, each 5\,ks long,
starting on  2011 July 13. 
This strategy was devised in order to cover the light curve in different 
orbital phases while maintaining the observing time per day 
reasonably short, thus not hindering gamma-ray burst (GRB) observations. 
However, most observations were cut short by the occurrence of GRBs 
and the discovery of a new soft gamma-ray repeater 
(starting from obsID 00031929005, see Table~\ref{igr16418:tab:xrtobs}). 
Observation 00032037001 was performed to determine the best 
offset from the intended target in order to alleviate the problem 
of single-reflection rings from the nearby source GX~340$+$0   
(these rings marginally affect the source while in the faint states, 
so that the analysis of the earlier segments was still feasible, 
with an appropriate choice of extraction regions)   
A new pointing (hence a new obsID) was adopted afterwards. 
In our work we also include observation 00043105001 
that serendipitously covers the region of \src. 
The 2011 July campaign lasted 18 days divided in 12 observations  
for a total on-source exposure of $\sim$ 39 \,ks.

For our binary orbital period search and analysis 
we retrieved the BAT `orbit-by-orbit' light curves (averaged over \sw's
orbital period of $\sim 90$\,m, 15--50\,keV) 
covering the data range from 2005 February 12 
to 2011 July 12 (MJD range 53413--55754) from the 
BAT Transient Monitor \citep[][]{Krimm2006_atel_BTM,Krimm2008_HEAD_BTM} 
page\footnote{http://swift.gsfc.nasa.gov/docs/swift/results/transients/ }. 
These data were further screened to exclude bad quality points 
(quality flag 1 and 2) and referred to the solar system barycentre (SSB)
by using the task {\sc earth2sun}.

The XRT data were processed with standard procedures 
({\sc xrtpipeline} v0.12.6), filtering and screening criteria by using 
{\sc FTOOLS}. 
Both WT and PC events were considered. 
The selection of event grades was 0--2 and 0--12, 
for WT and PC data, respectively (\citealt{Burrows2005:XRTmn}). 
Source events were accumulated within a annular/circular region 
(depending on whether pile-up correction was required or not, respectively) 
with an outer radius of 20 pixels (1 pixel $\sim2.36$\arcsec); 
and background events were accumulated from source-free regions far from 
the contamination of the single-reflection rings from GX~340.0$+$0. 
For our timing analysis we converted the event arrival times to the 
Solar system barycentre with the task {\sc barycorr}.  
Light curves were created for several values of signal-to-noise ratio (SNR) and 
number of counts per bin; all were corrected for PSF losses, vignetting and 
background. 
For our spectral analysis, we extracted events in the same regions as 
those adopted for the light curve creation; 
ancillary response files were generated with {\sc xrtmkarf},
to account for different extraction regions, vignetting, and PSF corrections. 
We used the latest spectral redistribution matrices in CALDB (20110705).

The UV/Optical Telescope \citep[UVOT, ][]{Roming2005:UVOTmn}  
observed \src\ simultaneously with the XRT with the 
`Filter of the Day' (FoD), i.e.\ the filter chosen for all observations 
to be carried out during a specific day in order to minimize the filter 
wheel usage. 
The exception is observation 00032037007, which was requested in the
$u$ filter, as opposed to FoD, to attempt to obtain a detection in that band. 
The data analysis was performed using the uvotimsum and 
{\sc uvotsource} tasks included in the FTOOLS. The latter 
task calculates the magnitude through aperture photometry within
a circular region and applies specific corrections due to the detector
characteristics.

All quoted uncertainties are given at 90\,\% confidence level for 
one interesting parameter unless otherwise stated. 
The spectral indices are parameterized as $F_{\nu} \propto \nu^{-\alpha}$, 
where $F_{\nu}$ (erg cm$^{-2}$ s$^{-1}$ Hz$^{-1}$) is the 
flux density as a function of frequency $\nu$; 
we adopt $\Gamma = \alpha +1$ as the photon index, 
$N(E) \propto E^{-\Gamma}$ (ph cm$^{-2}$ s$^{-1}$ keV$^{-1}$).

\begin{figure}
\begin{center}
\includegraphics*[angle=270,width=8cm]{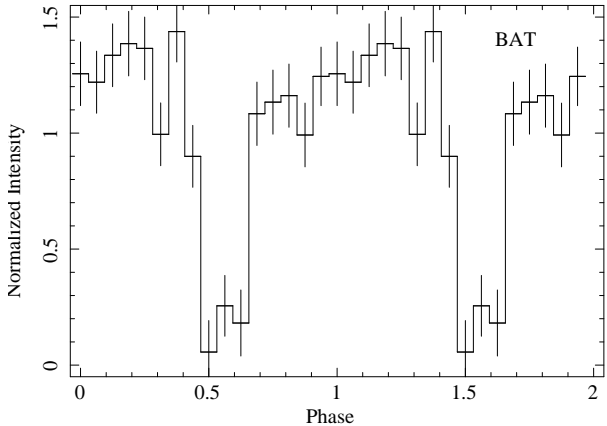}
\end{center}
\caption{ BAT light curve folded at 
$P_{\rm orb}=3.73886$\,d and  $T_{\rm epoch}=$ MJD 53560.20000. 
}
\label{igr16418:fig:batlcvphased}
\end{figure}

 	 \section{Results \label{igr16418:results} }

 	 \subsection{Orbital period from BAT data \label{igr16418:porb} }

For the orbital modulation, we considered the BAT data in the time range 
MJD 53414--55754. 
With a standard folding analysis of the Solar system barycentred light curves, 
we measured an orbital period of $3.740\pm0.002$ days (1-$\sigma$ error), 
which is fully consistent 
with the value obtained by \citet{Levine2010}, $P_{\rm orb}=3.73886\pm0.00028$\,d. 
Therefore, hereafter we shall adopt the more precise measure by \citet{Levine2010},
and we shall refer to $T_{\rm epoch}=$ MJD 53560.20000 \citep{Corbet2006:atel779}.  
Fig.~\ref{igr16418:fig:batlcvphased} shows the BAT light curve folded 
over this period. 

The BAT data show an eclipse centred at $\phi \sim 0.55$.  
The depth is consistent with a total eclipse, 
and the eclipse FWHM duration is $0.17\pm0.05$ of the orbital period.
The folded light curve is quite flat;
the peak-to-trough pulsed fraction,  
$PF_{\rm pt} = (F_{\rm max} - F_{\rm min})/(F_{\rm max} + F_{\rm min})$, 
where $F_{\rm max}$ and $F_{\rm min}$ are
the observed background-subtracted count rates at the peak and
at the minimum, is $78\pm12$\,\% (1-$\sigma$ error).

 	 \subsection{X-ray position  \label{igr16418:position} }

We used 20.4\,ks of PC mode data and simultaneous \sw/UVOT images to obtain 
an astrometrically-corrected position 
\citep[see ][]{Evans2009:xrtgrb_mn, Goad2007:xrtuvotpostions} of: 
RA(J$2000) = 250.46104$, Dec(J$2000) =-45.54092$, which is equivalent to: 
RA(J$2000)=16^{\rm h}\, 41^{\rm m}\, 50\fs65$, 
Dec(J$2000)=-45^{\circ}\, 32^{\prime}\, 27\farcs3$, 
with an uncertainty of 1\farcs9  (90\,\% confidence 
level)\footnote{This new UVOT-enhanced XRT position takes advantage of the new 
calibration of the XRT to UVOT detector co-ordinates of 2011 August \citep{Evans2011:reprocessedgcn}; 
see, also: http://www.swift.ac.uk/reprocessed.php. }. 
This position is $2\farcs5$ from the source 2MASS J16415078$-$4532253  
and $4\farcs3$ from the \xmm\ position \citep{Walter2006}, and it allows us to 
confirm the 2MASS source as the optical counterpart of \src.

\begin{figure}
\begin{center}
\includegraphics*[angle=270,width=8cm]{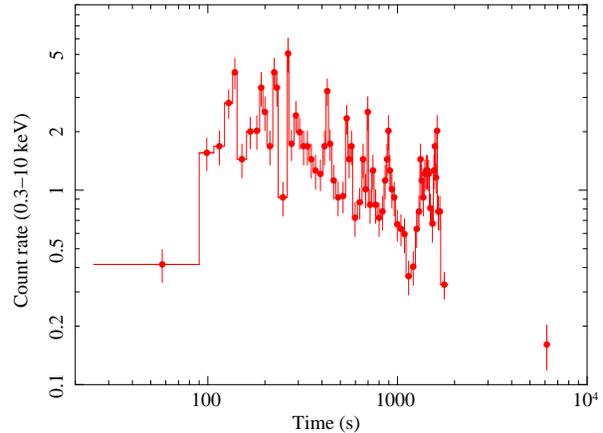}
\end{center}
\caption{\sw/XRT 0.3--10\,keV light curve of \src\ during observation 00031929001, 
background-subtracted, corrected for pile-up, PSF losses, and vignetting (SNR $=5$). 
Time is referred to the start of the observation. 
}
\label{igr16418:fig:xrtlcv001}
\end{figure}

 	 \subsection{UVOT optical counterpart \label{igr16418:uvot} }

At the XRT position of \src, no detection was achieved in any UV band 
down to a limit of $w1>21.78$,  $m2>21.66$, $w2>21.89$ mag, respectively. 
A marginal detection ($2.9 \sigma$) was obtained in the $u$ band, at 
$22.28\pm0.40$ mag. 
These magnitudes are on the UVOT photometric system described in 
\citet{Poole2008:UVOTmn}, and are not corrected for Galactic or intrinsic 
extinction.  
At this position, the column due to Galactic absorption 
is $1.59\times 10^{22}$\,cm$^{-2}$ \citep{LABS}. 
By adopting the conversion into $A_V$ \citep{avnh},  
$N_{\rm H} = 1.79 \times 10^{21} A_V $ cm$^{-2}$, this 
corresponds to \citep{CCM1989:ABS} $A_{u}=14.2$\,mag in the UVOT $u$ band.

\begin{figure*}
\begin{center}
\includegraphics*[angle=270,width=17.5cm]{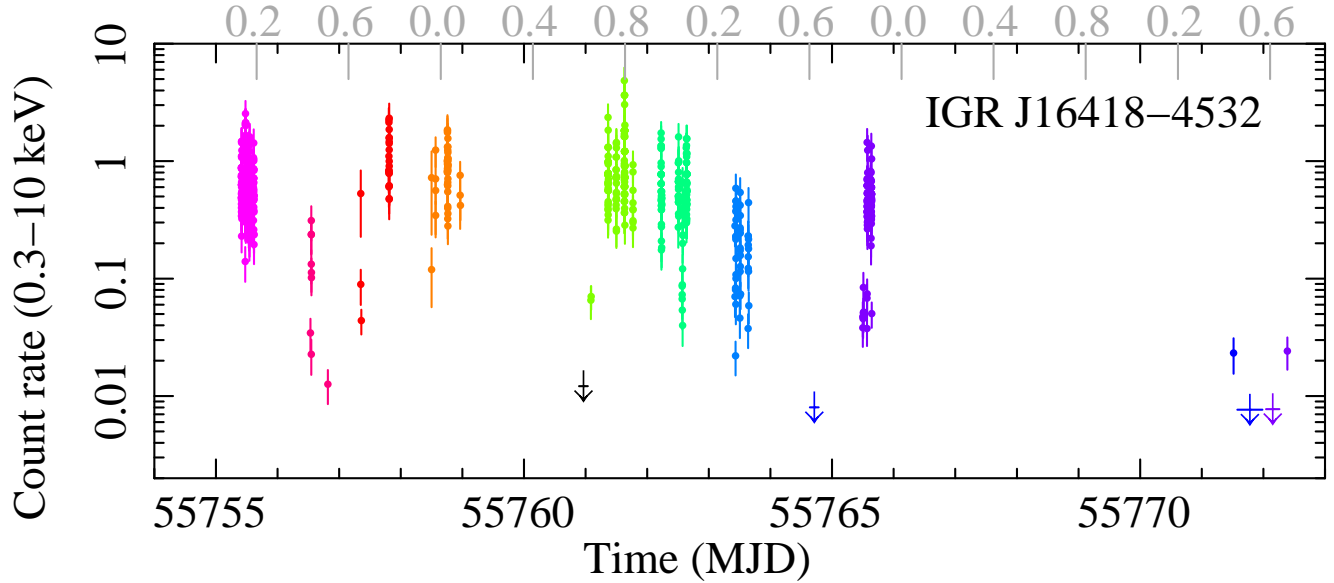}
\end{center}
\caption{\sw/XRT 0.3--10\,keV light curve of \src\ during the 2011 July monitoring 
program, background-subtracted and corrected for pile-up, PSF losses, and vignetting 
(SNR$=3$).    Downward-pointing arrows are 3$\sigma$ upper limits. 
Different colours (see the electronic edition) mark different observations 
(see Table~\ref{igr16418:tab:xrtobs}). The top axis reports the phase 
(with $P_{\rm orb}=3.73886$\,d and $T_{\rm epoch}=$ MJD 53560.20000).  
}
\label{igr16418:fig:xrtlcv_mjd}
\end{figure*}

\begin{figure*}
\begin{center}
\includegraphics*[angle=270,width=17.5cm]{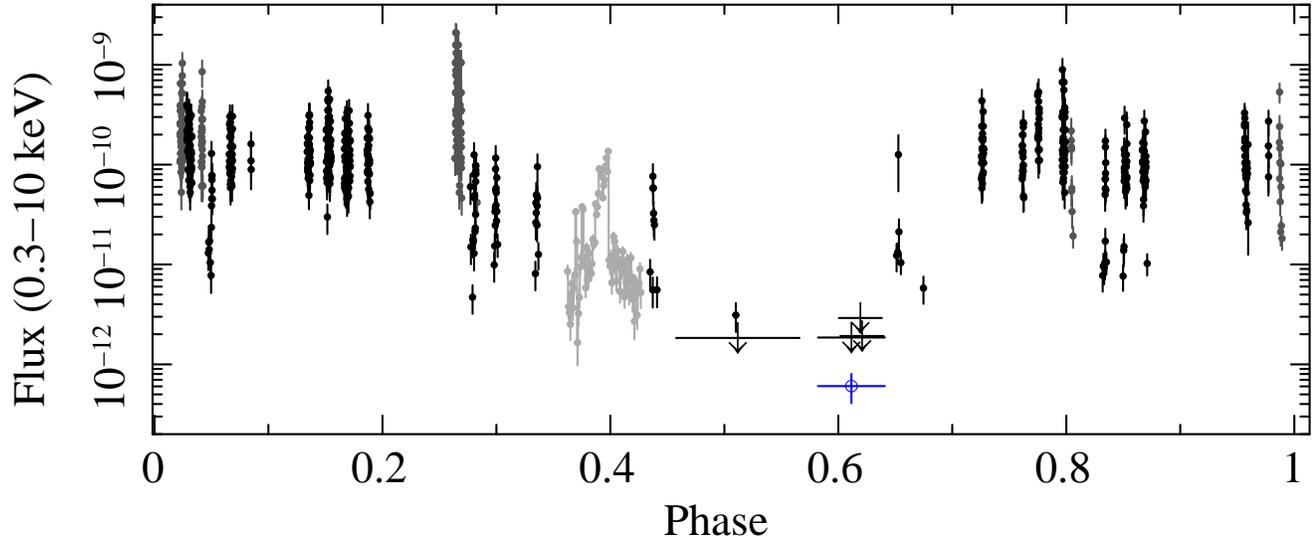}
\end{center}
\caption{\sw/XRT 0.3--10\,keV flux light curve of \src, 
folded at $P_{\rm orb}=3.73886$\,d and $T_{\rm epoch}=$ MJD 53560.20000. 
The data were collected in 2011 February and June (dark grey filled circles),
and 2011 July (black filled circles). 
Downward-pointing arrows are 3$\sigma$ upper limits. 
The open blue circle is a detection obtained by combining three observations 
(00032037001, 00032037004, and snapshots 1--3 of 00032037007). 
Also included are the 2004 August XMM data (light grey 
curve).  
}
\label{igr16418:fig:xrtlcv_phase}
\end{figure*}

 	 \subsection{Spin period  search \label{igr16418:pspin} }

We looked for evidence of the $1246\pm100$ s spin periodicity reported by 
\citet{Walter2006} in the Solar system barycentred XRT light curves. 
We only considered observations where the source was significantly detected 
(thus excluding obsIDs 00032037001 and 00032037004).

We searched each single observation longer than one spin period 
and the total combined dataset within $\pm300$\,s (corresponding to a $\pm3\sigma$ 
error reported by \citealt{Walter2006}) 
by applying both a Fourier transform and a $Z^2_n$ analysis 
(a method that does not require the binning of the data; \citealt{Buccheri1983}). 
No statistically significant signal was detected either in the individual 
observations or in the combined dataset. With respect to the latter, the upper 
limit on the pulsed fraction, (defined as the semi-amplitude of the sinusoidal 
modulation divided by the mean count rate), computed according to 
\citet{Vaughan1994}, is 42\,\% at the 99\,\% confidence level and for periods 
between 946 and 1546\,s.

This upper limit is only marginally compatible with the $(64\pm10)$\,\% value 
measured by \citet{Walter2006}. We note however that they were able to 
detect pulsations only during a fraction of their \emph{XMM-Newton} data 
and in particular while the source was not flaring, and that our data were 
collected in a variety of states of brightness and across almost all 
orbital phases. 
So our upper limit and their measure may not be straightforward to compare.

 	 \subsection{Light curves  \label{igr16418:lcvs} }

Fig.~\ref{igr16418:fig:xrtlcv001} shows the 0.3--10\,keV light curve of \src\ 
during the 2001 February 18 observation, binned in order to achieve a 
SNR of 5. The source reached $\sim 5$ counts s$^{-1}$ 
during the initial flare and then decayed through a series of smaller flares. 
The lowest point is at 0.16 counts s$^{-1}$. 
When binned at a SNR of 3 the dynamical range, during 
this observation, is of order 50. 

Fig.~\ref{igr16418:fig:xrtlcv_mjd} 
shows the 0.3--10\,keV light curve of \src\ of the whole 2011 July campaign 
after bin-by-bin background-subtraction and corrections for pile-up, PSF losses, and vignetting, 
binned at SNR $=3$. 
The light curve starts at phase 0.16 (assuming a period of $P_{\rm orb}=3.73886$\,d 
and an initial epoch $T_{\rm epoch}=$ MJD 53560.20000) 
and monitors the source state through over three full periods. 
Superimposed on the long-term orbital modulation, 
which follows that seen in the BAT data,
flaring is observed on short time scales, as shown in 
Fig.~\ref{igr16418:fig:xrtlcv001}.  
This behaviour has been observed in most SFXTs  
\citep[][]{Sidoli2008:sfxts_paperI,Romano2009:sfxts_paperV,Romano2011:sfxts_paperVI,Romano2010:sfxts_18483}. 
 
The lowest point (as observed by XRT) was collected during the eclipse, 
by combining three observations 
(00032037001, 00032037004, and 
snapshots\footnote{Part of an observation with a continuous viewing of the target, 
with typical timescales of $10^3$\,s. } 1--3 of 00032037007, see open blue point 
in Fig.~\ref{igr16418:fig:xrtlcv_phase}) 
at $2.5\times10^{-3}$ counts s$^{-1}$. This corresponds to 
an unabsorbed 0.3--10\,keV flux of $\sim6.0\times10^{-13}$ erg cm$^{-2}$ s$^{-1}$,
and to a luminosity of 1.2$\times10^{34}$ erg s$^{-1}$ 
(assuming the optical counterpart distance of 13\,kpc). 
The lowest point during the campaign {\it outside} the eclipse was recorded on 
MJD 55763.43 at 0.022 counts s$^{-1}$, corresponding to 
an unabsorbed 0.3--10\,keV flux of $\sim4.7\times10^{-12}$ erg cm$^{-2}$ s$^{-1}$,
and to a luminosity of 9.5$\times10^{34}$ erg s$^{-1}$. 
As the peak count rate is reached on MJD 55610.08 at $\ga 8$ counts s$^{-1}$ 
the observed dynamical range of this source is at least 370,
(1400 considering the points within the eclipse).

\begin{figure}
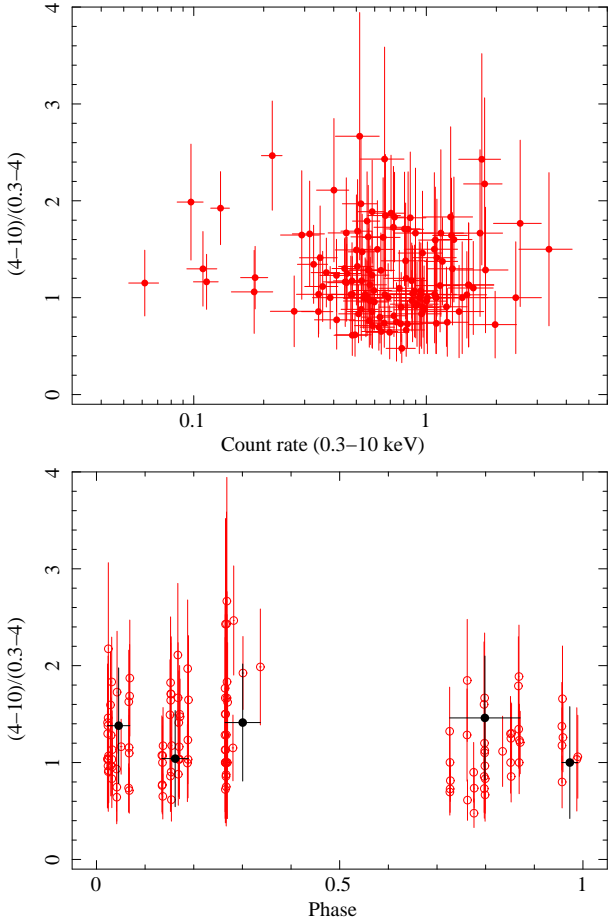

\begin{center}
\includegraphics*[angle=270,width=8cm]{figure5a.ps}
\includegraphics*[angle=270,width=8cm]{figure5b.ps}
\end{center}
\caption{\sw/XRT hardness ratio (red open circles, 
30 counts per bin)
as a function of count rate (top) and phase (bottom).
The black points are weighted means. 
}
\label{igr16418:fig:xrtlcv_hr}
\end{figure}

Fig.~\ref{igr16418:fig:xrtlcv_phase} shows the 0.3--10\,keV 
light curve, folded using 
$P_{\rm orb}=3.73886$\,d and $T_{\rm epoch}=$ MJD 53560.20000. 
We adopted count rate to flux conversion factors derived from spectral 
fits of each observation (Section~\ref{igr16418:spectra}) 
and subsequent interpolation across observations where fitting was not feasible. 
We also include the $\sim 10$\,ks 2004 August 19 \xmm/EPIC-pn observation \citep{Walter2006}, 
binned at 256\,s, and converted into flux by adopting a single conversion factor
obtained from fitting the mean spectrum with an absorbed power-law model 
($\Gamma=1.40\pm0.05$, $N_{\rm H}=(10.5\pm0.5)\times10^{22}$ cm$^{-2}$, 1-$\sigma$ errors; 
$\chi^2=$0.99/169 dof). 
The eclipse lasts $\Delta \phi \sim 0.2$ and is centred at $\phi \sim 0.55$,
as observed in the BAT data (Fig.~\ref{igr16418:fig:batlcvphased}).

Fig.~\ref{igr16418:fig:xrtlcv_hr} shows the 
hardness ratio (HR defined as H/S, where H is the count rate in the
4--10 keV band and S is the count rate in the 0.3--4 keV band),  
as a function of count rate (top), and orbital phase (bottom),
where the light curve was binned to have 30 counts per bin (red open circles).
A fit to a constant value yields a hardness ratio value of $1.01\pm0.03$ 
and $\chi^2_{\nu}=0.85$ for 130 degrees of freedom (dof), so that the HR  
is shown to be roughly constant over two orders of magnitude in count rate.
Furthermore, the HR were also weighted across several phase intervals 
(black points); 
we cannot find evidence of variations of the HR as a function of phase.

 	 \subsection{Spectral analysis \label{igr16418:spectra} }

We first extracted PC spectra for each segment in which a detection was obtained
and a minimum of $\sim 100$ source counts were available. We did not consider the
WT data for our spectroscopic study as they were statistically poor and were heavily contaminated 
by the background from the single-reflection rings from GX~340.0$+$0. 
The data were rebinned with a minimum of 20 counts per energy bin to allow $\chi^2$ 
fitting, with the exception of low-count-statistics spectra, in which Cash 
\citep{Cash1979} statistics and spectra binned to 1 count per bin were used, instead.  
The spectra were fit in the 0.3--10\,keV energy range with a single absorbed 
power law model. More complex models were not required 
by the data. 
The results are reported in Table~\ref{igr16418:tab:xrtspec}. 
We note that the spectral parameters do not show significant variability 
within the large uncertainties, despite the large observed variations in flux 
throughout the campaign. 

Observation 00031929001 (shown as the dark grey filled circles at phase 0.27 in 
Fig.~\ref{igr16418:fig:xrtlcv_phase}, and detailed in 
Fig.~\ref{igr16418:fig:xrtlcv001}), 
in particular, deserves special attention.
Although it was performed as a follow-up of a false trigger (later attributed to 
the nearby GX~340$+$0), it nonetheless found the source in a very active state,
the brightest of the whole campaign, with its average 2--10\,keV unabsorbed 
flux of $\sim 2\times 10^{-10}$ erg cm$^{-2}$ s$^{-1}$,
corresponding to $\sim 4\times10^{36}$ \,erg\,s$^{-1}$ at 13\,kpc).
The peak flux was $\sim 2\times10^{-9}$\,erg cm$^{-2}$ s$^{-1}$
($\sim 4\times10^{37}$ \,erg\,s$^{-1}$). 
An absorbed power-law model yielded a high absorbing column  
$N_{\rm H}=(6.8_{-1.8}^{+2.3})\times 10^{22}$ cm$^{-2}$ 
(in excess of the Galactic one, $1.59\times 10^{22}$ cm$^{-2}$) 
and a photon index $\Gamma=1.4_{-0.5}^{+0.6}$ 
(see Table~\ref{igr16418:tab:xrtspec}).

 \begin{table} 	
\tabcolsep 3pt   
 \begin{center} 	
 \caption{XRT spectroscopy of \src. \label{igr16418:tab:xrtspec}} 	
 \label{} 	
 \begin{tabular}{lcccrr} 
 \hline 
 \noalign{\smallskip} 
 ObsID      &$N_{\rm H}$  &$\Gamma$  &Flux$^a$     &L$^{b}$ &$\chi^{2}_{\nu}/$dof    \\
\noalign{\smallskip} 
           & (10$^{22}$~cm$^{-2}$) &           &            &  & C-stat (dof)$^{c}$  \\
 \hline 
00031929001     &$6.77_{-1.78}^{+2.32}$ &$1.38_{-0.54}^{+0.61}$ &$21.36_{-2.44}^{+3.14}$& 43 &$0.532/18$  \\
00031929002     &$4.17_{-1.24}^{+1.70}$ &$1.04_{-0.45}^{+0.51}$ &$11.58_{-1.30}^{+1.39}$& 23 &$0.845/17$   \\
00043105001     &$5.28_{-2.54}^{+3.47}$ &$1.35_{-0.87}^{+0.99}$ &$ 3.85_{-0.83}^{+1.16}$&  8 &70.05(76)     \\
00031929003     &$5.20_{-0.82}^{+1.00}$ &$1.35_{-0.24}^{+0.27}$ &$ 8.27_{-0.58}^{+0.65}$& 17 &$0.935/58$   \\
00031929004     &$5.16_{-2.81}^{+4.17}$ &$0.09_{-0.75}^{+0.87}$ &$ 1.05_{-0.17}^{+0.21}$&  2 &169.46(141)  \\
00031929007     &$3.93_{-0.97}^{+1.24}$ &$1.04_{-0.36}^{+0.40}$ &$ 7.62_{-0.69}^{+0.72}$& 15 &$0.647/28$   \\
00032037002     &$4.59_{-1.00}^{+1.23}$ &$1.05_{-0.34}^{+0.38}$ &$ 6.32_{-0.54}^{+0.56}$& 13 &$0.831/31$   \\
00032037003     &$5.65_{-1.96}^{+2.80}$ &$0.62_{-0.55}^{+0.62}$ &$ 2.51_{-0.31}^{+0.32}$&  5 &$1.114/16$   \\
00032037005     &$5.14_{-1.39}^{+1.72}$ &$1.01_{-0.40}^{+0.45}$ &$ 3.88_{-0.37}^{+0.40}$&  8 & $0.838/27$ \\
\noalign{\smallskip} 
  \hline
  \end{tabular}
  \end{center}
  \begin{list}{}{} 
  \item[$^{\mathrm{a}}$ Unabsorbed 2--10\,keV fluxes ($10^{-11}$ erg~cm$^{-2}$~s$^{-1}$). ] 
  \item[$^{\mathrm{b}}$ 2--10\,keV luminosities in units of $10^{35}$ erg~s$^{-1}$, at 13~kpc.] 
  \item[$^{\mathrm{c}}$ Cash statistics (C-stat).] 
  \end{list} 
  \end{table}

 	 \section{Discussion \label{igr16418:discussion}}

%
%
%
\src\ is a candidate SFXT, based on the hard X-ray behaviour \citep{Sguera2006}, 
for which we know both the orbital 
\citep[$P_{\rm orbit}\sim 3.7$\,d,][]{Corbet2006:atel779} 
and the spin  \citep[$\sim 1250$\,s,][]{Walter2006} periods. 
These values place \src\ in the region of the Corbet diagram \citep{Corbet1986} 
of the wind-accreting supergiant high-mass X-ray binaries. 
In this Paper we have taken advantage of the short orbital period which makes \src\ 
a very good source to monitor with the XRT, to study the X-ray properties of this source
as a function of the orbital phase, in particular, to address the issue of its SFXT nature.

The \sw/BAT and XRT folded light curves of 
\src\ (Fig.~\ref{igr16418:fig:batlcvphased}, and Fig.~\ref{igr16418:fig:xrtlcv_phase}) show 
the presence of a dip that is consistent with zero intensity. 
When interpreted as an eclipse, it can be used to infer the nature of the 
primary of the binary system. 
The BAT lightcurve of \src\ (excluding the eclipse)
does not clearly show a flux modulation with the orbital 
phase, suggesting that the eccentricity is not large. 
This is in agreement with the observation that in many X-ray binary 
systems with nearly circular orbits and short orbital periods 
($\leq$\,4\,days), the eccentric orbit produced by the explosion
of the supernova is circularized on timescale of $10^6$\,yr 
(consistent with the evolution timescale of an O star),
due to viscous tidal interactions \citep[see e.g.][]{Press1975}.   
Thus, for simplicity, we assume a circular orbit 
($P_{\rm orb}= 3.73886$\,d). 
Let us adopt the typical values of stellar masses and radii \citep{Martins2005} 
for a primary of spectral type O8.5 and luminosity classes I, III, or V, 
and for the neutron star a mass of 
$M_{\rm NS}=1.4$\,M$_\odot$, and a radius of $R_{\rm NS}=10$\,km. 
By combining Kepler's third law with the
relationship given by \citet{Rappaport1983} for the radius of the 
primary $R$, 
\begin{equation} \label{rappa}
(R/a)^2= \cos^2 i +\sin^2 i \, \sin^2 \theta_{\rm e},
\end{equation}
where $a$ is the semi-major axis of the orbit, $i$ is the inclination,
and $\theta_{\rm e}$ is the eclipse semi-angle, we find that the 
observed $\Delta\phi=0.2$ is inconsistent with a luminosity class III or V primary.
We therefore conclude that indeed the primary is an O8.5I supergiant star
(with mass, radius, luminosity, effective temperature given by \citet{Martins2005},
$M_{\rm OB}=31.5$\,M$_\odot$, $R_{\rm OB}=21.4$\,R$_\odot$, $\log L_{\rm OB}/L_\odot=5.65$, 
$T_{\rm eff} = 32274$\,K). 
Therefore, the distance to \src\ is $\sim 13$\,kpc \citep{Rahoui2008}.

During the eclipse of \src, 
we detect a non-zero flux with the XRT
($L_{\rm X} \sim 3 \times 10^{34}$\,erg\,s$^{-1}$,  
see Fig.~\ref{igr16418:fig:xrtlcv_phase}).
Detections of HMXBs during the eclipse are quite common 
\citep[see e.g.\ ][]{vanderMeer2005,Clark1988,Sako1999}
and can be ascribed to X-ray photons 
scattered by the stellar wind of the OB star
and discrete fluorescent and recombination lines
from highly ionized gas around the compact object.
The decrease in X-ray flux, when the compact object
is approaching the eclipse,
is related to the increase of the 
column density along the line of sight,
through the stellar wind produced by the OB star.
From observations of other wind-fed X-ray pulsars,
such as 4U~1700$-$377, Vela X-1, OAO~1657$-$415
\citep[see e.g.\ ][]{vanderMeer2005,Sako1999,Audley2006}
the absorption column density during the eclipse
ranges from $\approx 10^{23}$\,cm$^{-2}$
up to $\approx 10^{24}$\,cm$^{-2}$.

We can compare the X-ray luminosity and absorption column density 
of \src\ during the eclipse
with those calculated by \citet{vanderMeer2005} 
for 4U~1700$-$377,
a wind-fed HMXBs with properties similar to \src\ 
(orbital period $P_{orb} = 3.4$\,d,
companion star of spectral type O6.5Iaf).
\citet{vanderMeer2005} observed 4U~1700$-$377 
with XMM-Newton during an eclipse; they found an
X-ray luminosity of $\approx 7 \times 10^{34}$\,erg\,s$^{-1}$ 
(in agreement with the X-ray luminosity of \src\ during the eclipse)
and a column density of $\approx 10^{24}$\,cm$^{-2}$.

The XRT light curve (Figs.~\ref{igr16418:fig:xrtlcv001} and ~\ref{igr16418:fig:xrtlcv_mjd})
shows a large dynamic range, up to 1--2 orders of magnitude within a single 
snapshot  
and about 3 orders of magnitude overall (even when the low-intensity state is excluded). 
While this dynamic range is lower than that observed in the classical
 \citep{Chaty2010arXiv1012.2318C} SFXTs, which reach 4--5 orders of magnitude 
in dynamic range (e.g.\ IGR~J17544$-$2619),
it is indeed typical of intermediate SFXTs (e.g., IGR~J18483$-$0311). 
The 2011 February flare  (Fig.~\ref{igr16418:fig:xrtlcv001} and dark grey filled 
circles at phase 0.27 in Fig.~\ref{igr16418:fig:xrtlcv_phase}), in particular, 
reached $\sim 2\times10^{-9}$\,erg cm$^{-2}$ s$^{-1}$ (2--10\,keV, unabsorbed), 
corresponding to a luminosity of $\sim 4\times10^{37}$ \,erg\,s$^{-1}$,   
which is typical of SFXT outbursts as we have observed them with \sw/XRT 
\citep[e.g., ][]{Romano2008:sfxts_paperII,Romano2009:sfxts_paper08408,
Sidoli2009:sfxts_paperIII,Sidoli2009:sfxts_sax1818}.

%
\begin{figure}
\begin{center}

\hspace{-0.5truecm}
\includegraphics[width=8.7cm]{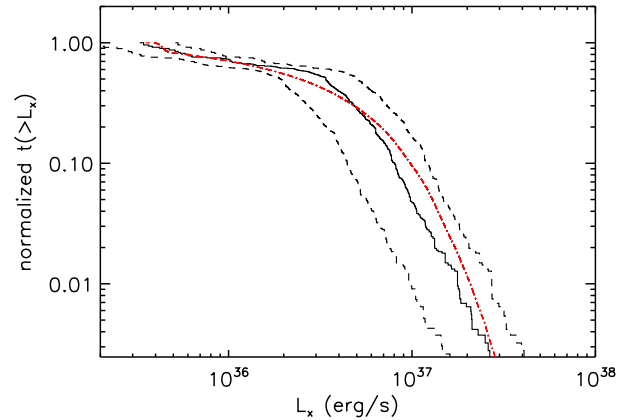} 
\vspace{-0.5truecm}
\end{center}
\caption{Observed cumulative luminosity distribution (solid black line) and relative
uncertainties (dashed black lines), compared with a possible solution 
(red dot-dashed red line; see text).
}
\label{igr16418:fig:cumulative-distrib}
\end{figure}
%
\begin{table}
\begin{center}
\caption{Wind parameters values for \src. \label{igr16418:tab:wind-parameters}}
\begin{tabular}{lcc}
\hline
Parameter  & min. value & max. value \\
\hline
$\dot{M}_{\rm tot}$ & $4 \times 10^{-7}$\,M$_\odot$\,yr$^{-1}$ & $8 \times 10^{-7}$\,M$_\odot$\,yr$^{-1}$ \\
$f$        & $0.85$  & $0.95$ \\
$v_\infty$  & $800$\,km\,s$^{-1}$ & $1300$\,km\,s$^{-1}$ \\
$\beta$    & $0.8$ & $1.2$ \\
$\zeta$    &  $0.8$ & $1.6$ \\
$\gamma$   & $-1.5$ & $-0.5$ \\
$M_{\rm a}$ & $5 \times 10^{16}$\,g & $5 \times 10^{19}$\,g \\
$M_{\rm b}$ & $5 \times 10^{19}$\,g & $10^{21}$\,g \\
\hline
\end{tabular}
\end{center}
\end{table}
%
%
%
%
On the other hand, while the typical SFXT light curve is characterized by a faint
flaring activity with luminosity of $10^{33}-10^{34}$\,erg\,s$^{-1}$
\citep{Sidoli2008:sfxts_paperI,Romano2009:sfxts_paperV,Romano2011:sfxts_paperVI},
sporadically interrupted by flares with duration of a few hours and peak
luminosity of $10^{36}-10^{37}$\,erg\,s$^{-1}$, 
\src\ exhibits a flaring activity ranging from 
$\sim 10^{35}$\,erg\,s$^{-1}$ up to $\sim 4 \times 10^{37}$\,erg\,s$^{-1}$ 
(also see \citealt{Walter2006} and the light grey curve in 
Fig.~\ref{igr16418:fig:xrtlcv_phase}). 

To reproduce the observed X-ray variability, we applied the clumpy wind model 
of \citet{Ducci2009}, which we briefly summarize here. 
This model assumes that the wind of OB supergiants
is inhomogeneous, composed of dense clumps surrounded by a hotter
and homogeneous wind. Clumps follow a power-law mass distribution
\begin{equation} \label{Npunto}
p(M_{\rm cl})=k \left ( \frac{M_{\rm cl}}{M_{\rm a}} \right)^{-\zeta},
\end{equation}
where $k=(1 - \zeta)\,M_{\rm a}^{-\zeta}/(M_{\rm b}^{1-\zeta}-M_{\rm a}^{1-\zeta})$ 
is the normalization constant,
$M_{\rm cl}$ is the mass of a clump,
and $M_{\rm a}$ and $M_{\rm b}$ define the extrema of the mass range.
Assuming a spherical geometry for the clumps,
the initial clump dimension distribution is given by 
$\dot{N}_{M_{cl}} \propto R_{\rm cl}^{\gamma}$\,clumps\,s$^{-1}$,
where $R_{\rm cl}$ is the radius of the clump.
The clump mass loss rate $\dot{M}_{\rm cl}$ is related to the total
mass loss rate $\dot{M}_{\rm tot}$ by means of 
$f=\dot{M}_{\rm cl} / \dot{M}_{\rm tot}$,
the fraction of wind mass in the form of clumps.
Clumps and the homogeneous wind follow the $\beta$\emph{-velocity} law
\begin{math} \label{legge_velocita}
v(r) = v_{\infty}\left (1 - 0.9983\frac{R_{\rm OB}}{r} \right )^{\beta}
\end{math}
\citep{Castor1975},
where $\beta$ is a constant in the range $\sim 0.5-1.5$,
$R_{\rm OB}$ is the radius of the supergiant and 
$v_{\infty}$ is the terminal wind velocity.

\src\ was intensely monitored by \emph{Swift}/XRT 
for two orbital cycles (MJD 55755--55766). Each snapshot 
has a duration of $\sim 1$\,ks, for a total of up to $\sim 5$\,ks per day,
which is the typical timescale for the 
flare durations observed in SFXTs \citep{Sguera2006}.
For this reason, we cannot establish the number of observed flares and their duration,
therefore we cannot use this information
to compare the observed lightcurve with that calculated with
the clumpy wind model of \citet{Ducci2009}, as  
has been done in previous work 
\citep{Ducci2009,Romano2010:sfxts_18483,Ducci2010}. 
However, it is possible to compare the observed cumulative 
luminosity distribution out of eclipse
with that calculated with the clumpy wind model.
For the former, 
the \emph{Swift}/XRT count rate, measured in the $0.3$--$10$\,keV,
was converted to the $0.1$--$100$\,keV luminosity using the 
average spectral parameters obtained by \citet{Walter2006}
and \citet{Ducci2010}, and assuming a distance of 13\,kpc.
For the latter, we adopted the physical parameters for the O8.5I 
primary and the neutron star secondary reported above, and 
the force multiplier parameters \citep[see][]{Ducci2009}
obtained by \citet{Shimada1994} for a O8.5I star 
($k=0.375$, $\alpha=0.522$, $\delta=0.099$).

Fig.~\ref{igr16418:fig:cumulative-distrib} shows the 
\sw/XRT cumulative luminosity distribution (black solid line) 
with its 90\,\% uncertainties (black dashed lines). 
Fig.~\ref{igr16418:fig:cumulative-distrib} also shows the 
cumulative luminosity distribution (red dot-dashed line) 
calculated with the clumpy wind model by 
adopting the following parameter values:
mass loss rate $\dot{M}_{\rm tot}=6 \times 10^{-7}$\,M$_\odot$\,yr$^{-1}$, 
terminal velocity $v_\infty=1000$\,km\,s$^{-1}$, 
$\beta=1$, fraction of mass lost in clumps $f=0.9$, 
mass distribution power-law index $\zeta = 1.1$,  
power-law index of the initial clump dimension distribution $\gamma =-1$,
minimum clump mass $M_{\rm a}=5 \times 10^{18}$\,g and 
maximum clump mass $M_{\rm b}=10^{20}$\,g.
Further acceptable solutions can be found assuming the wind parameters in the
ranges reported in Table~\ref{igr16418:tab:wind-parameters}.

The wind parameters obtained for \src\ are roughly 
in agreement with those of HMXBs previously studied 
\citep[see ][]{Ducci2009,Romano2010:sfxts_18483,Ducci2010},
with the exception of $f$ and $v_\infty$.
The obtained low value of the terminal velocity
can be explained as follows: when the wind is highly ionized by the X-ray photons
emitted by the neutron star,
the high ionization modifies the dynamics of the line-driven stellar wind 
of OB supergiants (especially in close binary systems),
leading to a reduction of the wind velocity in the direction of the neutron star
(see e.g. \citealt{Stevens1990} and references therein).
The mass loss rate is lower than the typical
mass loss rate of O8.5I stars, which is of the order of 
$\approx 2 \times 10^{-6}$\,M$_\odot$\,yr$^{-1}$ \citep{Vink2000},
but in agreement with the hypothesis that 
the mass loss rates derived from homogeneous-wind model
measurements with the H$\alpha$ method are overestimated
by a factor of 2--10 if the wind is clumpy (see e.g. \citealt{Lepine2008}).
However, we cannot exclude the possibility that the X-ray variability
of \src\ is not totally due to the accretion
of an inhomogeneous wind. In fact, other mechanisms could be at work,
like e.g.\ transient accretion discs \citep{Taam1988} and 
intermittent accretion flow onto a neutron star \citep{Lamb1977,Ducci2010}.

Our results, based on our intense monitoring of \src, 
show that its X-ray properties (dynamical range, 
flaring activity and peak luminosity) are consistent with an 
intermediate SFXT nature.

\section*{Acknowledgments}

We thank the {\it Swift} team duty scientists and science planners. 
We also thank the remainder of the {\it Swift} XRT and BAT teams, S.\ Barthelmy 
and J.A.\ Nousek, 
in particular, for their invaluable help and support of the SFXT project as a whole. 
We thank A.\ Cucchiara for helpful discussions. 
We acknowledge financial contribution from the agreement ASI-INAF I/009/10/0.
This work was supported at PSU by NASA contract NAS5-00136. 
PE acknowledges financial support from the Autonomous Region of Sardinia 
through a research grant under the program PO Sardegna FSE 2007--2013, L.R. 7/2007 
``Promoting scientific research and innovation technology in Sardinia''.
We also thank the anonymous referee for swift comments that helped improve the paper.

\bsp

\label{lastpage}

\end{document}